\documentclass{PoS}
\pdfoutput=1
\usepackage[latin1]{inputenc}                        
\usepackage{amsmath}
\usepackage{bm}
\usepackage{mathtools}
\usepackage{xspace}
\usepackage{graphicx}
\usepackage{subfig}

\newcommand{\Ps} {\ensuremath{s}\xspace} 
\newcommand{\Pb} {\ensuremath{b}\xspace}
\newcommand{\Pf} {\ensuremath{f}\xspace}
\newcommand{\PB} {\ensuremath{B}\xspace} 
\newcommand{\PD} {\ensuremath{D}\xspace} 
\newcommand{\CP}{{\ensuremath{C\!P}}\xspace}   
\newcommand{\lhcb}{\mbox{LHCb}\xspace}
\newcommand{\dzero} {\mbox{D0}\xspace}
\newcommand{\bquark}{{\ensuremath{\Pb}}\xspace}
\newcommand{\bbarquark}{{\ensuremath{\kern 0.18em\overline{\kern -0.18em \Pb}{}}}\xspace}
\newcommand{\squark}{{\ensuremath{\Ps}}\xspace}
\newcommand{\B}{{\ensuremath{\PB}}\xspace}
\newcommand{\D}{{\ensuremath{\PD}}\xspace}
\newcommand{\Bz}{{\ensuremath{\B^0}}\xspace}
\newcommand{\Dnul}{{\ensuremath{\D^0}}\xspace}
\newcommand{\Bs}{{\ensuremath{\B^0_\squark}}\xspace}
\newcommand{\invfb}{\ensuremath{\mbox{\,fb}^{-1}}\xspace}
\newcommand{\asl}{\ensuremath{a_{\mathrm{sl}}}\xspace}
\newcommand{\afs}{\ensuremath{a_{\mathrm{fs}}}\xspace}
\newcommand{\asld}{\ensuremath{\asl^d}\xspace}
\newcommand{\afsd}{\ensuremath{\afs^d}\xspace}
\newcommand{\asls}{\ensuremath{\asl^s}\xspace}
\newcommand{\afss}{\ensuremath{\afs^s}\xspace}
\newcommand{\Bbar}{{\ensuremath{\kern 0.18em\overline{\kern -0.18em \PB}{}}}\xspace}

\newcommand{\Dbar}{{\ensuremath{\kern 0.18em\overline{\kern -0.18em \PD}{}}}\xspace}
\newcommand{\Bq}{\ensuremath{\B_q^0}\xspace}
\newcommand{\Bbarq}{\ensuremath{\Bbar_q^0}\xspace}
\newcommand{\Dbarnul}{\ensuremath{\Dbar^0}\xspace}
\newcommand{\BH}{\ensuremath{\PB^0_H}\xspace}
\newcommand{\BL}{\ensuremath{\PB^0_L}\xspace}
\newcommand{\f}{{\ensuremath{\Pf}}\xspace}
\newcommand{\fbar}{{\ensuremath{\kern 0.18em\overline{\kern -0.18em \Pf}{}}}\xspace}
\newcommand{\ket}[1]{\ensuremath{|#1\rangle}}              

\title{Measuring Semileptonic Asymmetries in LHCb}

\ShortTitle{Measuring Semileptonic Asymmetries in LHCb}

\author{\speaker{Suzanne Klaver}%
         \thanks{On behalf of the \lhcb Collaboration}\\
        The University of Manchester, Manchester, UK\\
        E-mail: \email{suzanne.klaver@cern.ch}}

\abstract{The \CP-violating flavour-specific asymmetry in neutral \bquark mesons provides a method for testing the Standard Model. The measurements from the \dzero experiment yield values of this asymmetry that disagree with the Standard Model at a level of 3.6 $\sigma$. This contribution discusses the latest \lhcb measurements in this sector both from \Bz mesons (\asld) and \Bs mesons (\asls). Using their 2011 dataset, corresponding to an integrated luminosity of 1.0 \invfb obtained in 2011, \lhcb measured a value of $\asls = (-0.06 \pm 0.50_{\text{stat}} \pm 0.36_{\text{syst}}) \%$. Combining the 2011 and 2012 datasets, with an integrated luminosity of 3 \invfb, \lhcb measured $\asld = (-0.02 \pm 0.19_{\text{stat}} \pm 0.30_{\text{syst}}) \%$. These are the most precise measurements of the parameters \asls and \asld to date. Plans for an updated result for \asls using the full 3 \invfb dataset are discussed. This will include new methods to determine detection asymmetries which are the dominating systematic uncertainty of the 2011 measurement.}

\FullConference{Flavorful Ways to New Physics\\
                 28-31 October 2014\\
                 Freudenstadt - Lauterbad, Germany}

\begin{document}

\section{Introduction to Neutral Meson Mixing and Semileptonic Asymmetries}
\noindent Neutral mesons can oscillate into their own antiparticle through a second order weak amplitude; this process is called mixing. Since these processes are heavily suppressed in the Standard Model, they are sensitive to new physics. The mixing of these neutral mesons, \Bq, can be described by the Schr{\"o}dinger equation:
\begin{equation}
i \frac{d}{dt}  \left( \! \begin{array}{c} \ket{\Bq (t) } \\ \ket{\Bbarq (t) } \end{array} \! \right) = \left( M - \frac{i}{2} \Gamma \right)  \left( \! \begin{array}{c} \ket{\Bq (t) } \\ \ket{\Bbarq (t) } \end{array} \! \right),
\label{eq:schroedinger}
\end{equation}
where $M$ and $\Gamma$ represent a $2\times2$ complex matrix that characterises the mixing. The neutral mesons relevant for this contribution contain a \bquark quark and are denoted by \Bq, where $q=\{d,s\}$. Their mass eigenstates can be described as a linear combination of the flavour eigenstates:
\begin{align}
\ket{\BL} &= p \ket{\Bq} + q \ket{\Bbarq}\\
\ket{\BH} &= p \ket{\Bq} - q \ket{\Bbarq},
\label{eq:masseigenstates}
\end{align}
where $L$ and $H$ indicate ``light'' and ``heavy'', and where $p$ and $q$ are complex coefficients satisfying $|p|^2 + |q|^2 = 1$. Mixing is characterised by the difference in mass, $\Delta m = m_H - m_L$, and decay width, $\Delta \Gamma = \Gamma_L - \Gamma_H$. 

There is \CP violation in mixing if $\mathcal{P}(\Bq \rightarrow \Bbarq) \neq \mathcal{P}(\Bbarq \rightarrow \Bq)$. By looking at flavour-specific decays, the amount of \CP violation in mixing can be measured. The final states of these decays indicate whether the \PB decayed as a \Bq or \Bbarq. One can measure the flavour-specific asymmetry $a_{\mathrm{fs}}$, which is defined as:
\begin{equation}
a_{\mathrm{fs}} = \frac{ \Gamma (\Bbarq \rightarrow \Bq \rightarrow \f) - \Gamma (\Bq \rightarrow \Bbarq \rightarrow \fbar) }{ \Gamma (\Bbarq \rightarrow \Bq \rightarrow \f) + \Gamma (\Bq \rightarrow \Bbarq \rightarrow \fbar) } = \frac{ 1-|q/p|^4 }{ 1+|q/p|^4 }.
\label{eq:afs}
\end{equation}
Hence, there is \CP violation in mixing when $|q/p| \neq 1$. The Standard Model predicts values of \afs that are zero compared to the experimental resolution: $\afsd = (-4.1 \pm 0.6) \times 10^{-4}$ for \Bz decays and $\afss = (1.9 \pm 0.3) \times 10^{-5}$ for \Bs decays \cite{lenzsm2}. Semileptonic decays are flavour specific, i.e. the charge of the lepton indicates the flavour of the \PB, and are therefore excellent for measuring this asymmetry which is then called the semileptonic asymmetry: \asl. The relevant semileptonic decays are $\Bs \rightarrow D_s^-  \ell^+ \nu_{\ell}$, $\Bz \rightarrow D^- \ell \nu_{\ell}$ and $\Bz \rightarrow D^{*-} \ell \nu_{\ell}$.

\subsection{Methods to measure \asl}
\noindent There are different methods to measure semileptonic asymmetries. The method used by the \dzero experiment measures the inclusive like-sign dilepton asymmetry. Like-charge lepton pairs can be produced by \bquark \bbarquark pairs where both \bquark hadrons decay semileptonically, but where one \bquark forms a neutral meson that oscillates before decaying. This asymmetry is time-integrated and is defined as:
\begin{equation}
a_{\text{sl}} = A_{\ell \ell} = \frac{ \Gamma(\ell^+\ell^+) - \Gamma(\ell^-\ell^-) }{ \Gamma(\ell^+\ell^+) + \Gamma(\ell^-\ell^-) }.
\label{eq:asldilepton}
\end{equation}
The value that is found using this method is a linear combination of \asls and \asld, which are the corresponding contributions from \Bs and \Bz decays respectively. The \dzero experiment has found a value for this of $\asl = (-0.79 \pm 0.17_{\text{stat}} \pm 0.09_{\text{syst}})\%$ \cite{d0new} which is 3.6 $\sigma$ away from the Standard Model prediction.

In other experiments, such as \lhcb, only one of the \bquark mesons is reconstructed and a time-dependent measurement of \asl is made using the following expression:
\begin{equation}
\frac{ N(B,t) - N(\bar{B},t) }{ N(B,t) + N(\bar{B},t) } = \frac{\asl}{2} \left[ 1 - \frac{\cos \Delta m t}{\cosh \frac{1}{2} \Delta \Gamma t} \right].
\label{eq:asl}
\end{equation}
For the measurement of \asld, the time-dependence is an important aspect of the analysis. The measured asymmetry also includes detection asymmetries, $A_D$, for the various final state particles, and production asymmetries, $A_P$, which are caused by the fact that the initial state only contains protons, and no anti-protons. Including these, the expression for the measured asymmetry in terms of \asld is:
\begin{equation}
A_{\text{meas}}(t) = \frac{ N(B,t) - N(\bar{B},t) }{ N(B,t) + N(\bar{B},t) } = \frac{\asld}{2} + A_D + \left[ \frac{\asld}{2} - A_P \right]  \frac{\cos \Delta m_d t}{\cosh \frac{1}{2} \Delta \Gamma_d t}.
\label{eq:asld}
\end{equation}

Integrating over decay time, and accounting for the detection efficiency as a function of decay time, $\varepsilon(t)$, one obtains the expression for the time-integrated measured asymmetry for \asls:
\begin{equation}
A_{\text{meas}} = \frac{N(\mu^+D_s^-) - N(\mu^- D_s^+)}{N(\mu^+D_s^-) + N(\mu^- D_s^+)} = \frac{\asls}{2} + A_D + \left( \frac{\asls}{2} - A_P \right) \frac{ \int_0^\infty e^{-\Gamma_s t} \cos(\Delta m_s t) \varepsilon(t) \, dt }{ \int_0^\infty e^{-\Gamma_s t} \cosh(\frac{1}{2} \Delta \Gamma_s t) \varepsilon(t) \, dt}.
\label{eq:asls}
\end{equation}

Since the value of $\Delta m_s$ is very large ($1.2 \times 10^{-8}$ MeV), \Bs oscillates fast and the integral yields a small value ($\sim 1\%$). Combined with the small values of \asls and $A_P$ ($\sim 1\%$), this is of the order $<10^{-4}$ which is negligible compared to the uncertainties on the measurement. The expression therefore reduces to:
\begin{equation}
A_{\text{meas}} = \frac{N(\mu^+D_s^-) - N(\mu^- D_s^+)}{N(\mu^+D_s^-) + N(\mu^- D_s^+)} = \frac{\asls}{2} + A_D.
\label{eq:aslsred}
\end{equation}

\begin{figure}
\centering
\begin{minipage}[b]{0.47\textwidth}
\includegraphics[width=\linewidth]{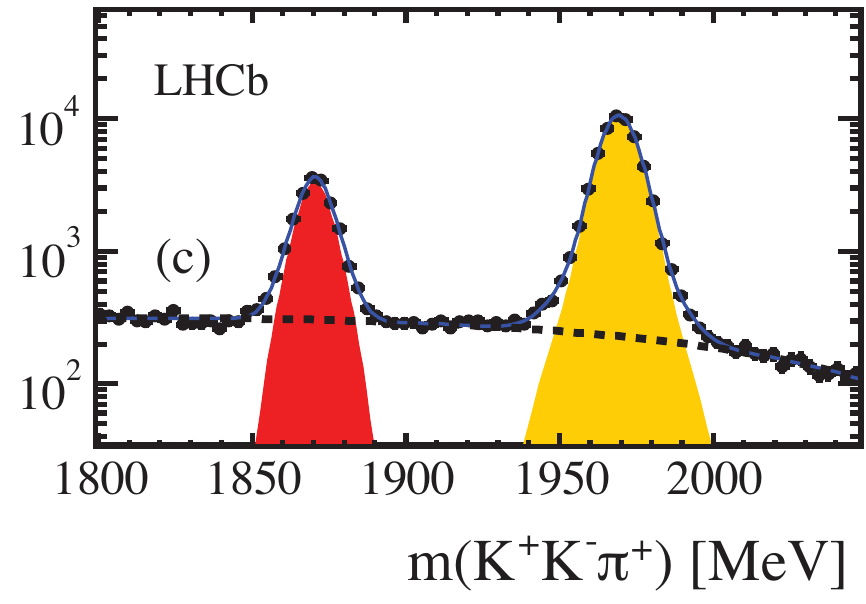}
\caption{Invariant mass distributions for $K^+K^-\pi^+$, where the $K^+K^-$ invariant mass is within 20 MeV of the $\phi$ meson mass.}
\label{fig:aslsresult}     
\end{minipage}
\quad
\hspace{0.03\linewidth}
\begin{minipage}[b]{.40\textwidth}
\includegraphics[width=\linewidth]{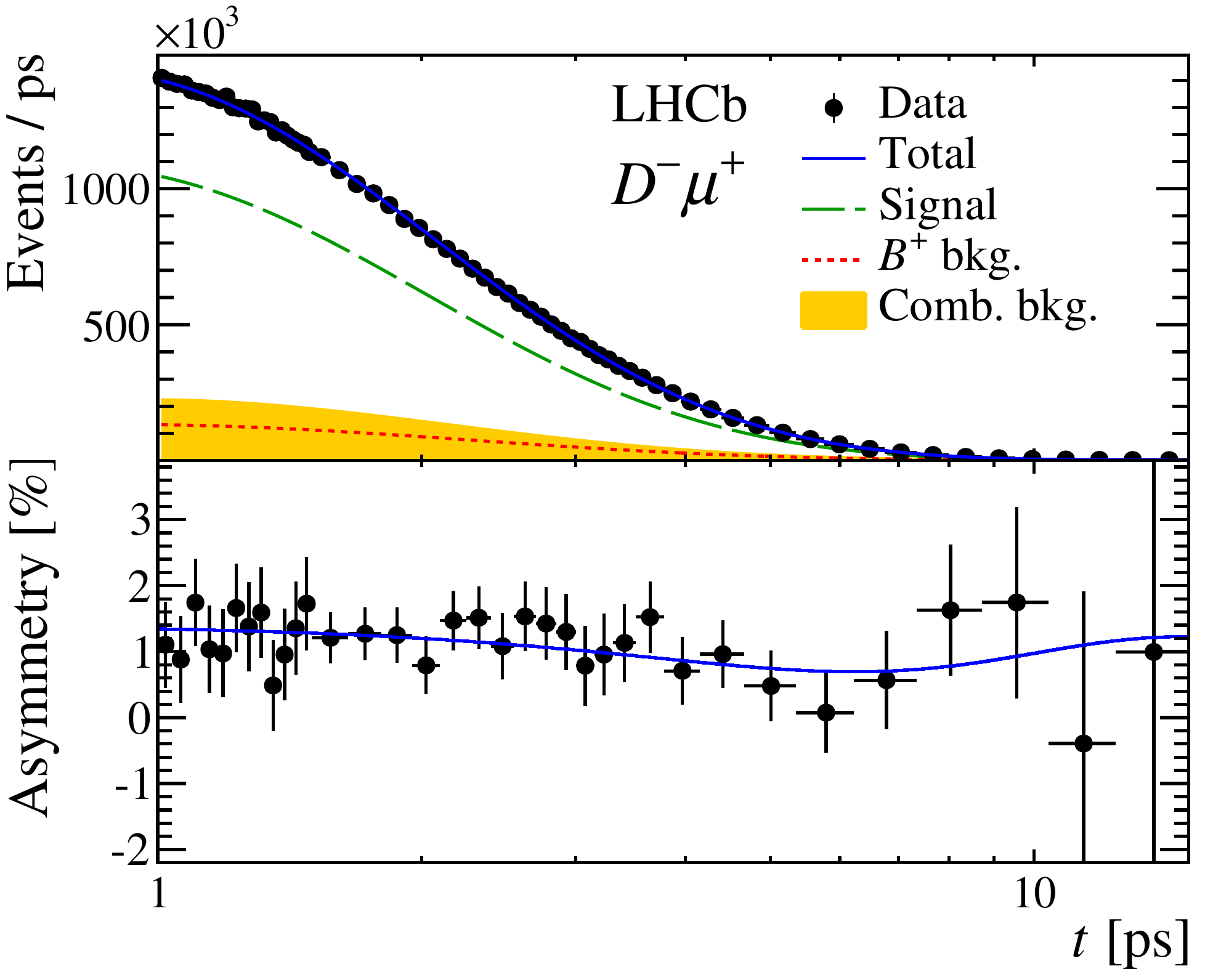}
\vspace{-7pt}
\caption{Decay rates and charge asymmetry versus decay time for the $D^- \mu^+$ sample used for the \asld analysis.}
\label{fig:asldresult}
\end{minipage}
\end{figure}
 
\section{Latest results on \asld and \asls from \lhcb}
\noindent The \lhcb detector has an excellent momentum resolution. The tracking system provides a measurement of momentum, $p$, of charged particles with a relative uncertainty that varies from 0.5\% at low momentum to 1.0\% at 200 GeV/$c$ \cite{lhcb-performance}. This allows for separation between \Bz and \Bs mesons. It also has a high muon identification efficiency which makes it very suitable for the \asl measurements. In 2011, \lhcb gathered an integrated luminosity $\mathcal{L} = 1.0 \invfb$ of data at a centre-of-mass energy of $\sqrt{s} = 7$ TeV. In 2012, $\mathcal{L} = 2 \invfb$ was gathered with $\sqrt{s} = 8$ TeV. More information on the \lhcb detector can be found in \cite{lhcb}.
 
The analysis of \asls using data from the \lhcb detector is based on data gathered in 2011. The current \asls analysis at \lhcb \cite{aslslhcb} studies the decay of $\Bs \rightarrow D_s^-  \mu^+ \nu_{\mu}$, where $D_s^-  \rightarrow \phi ( \rightarrow K^+ K^-) \pi^-$. The detection asymmetries for the final state particles hence come from the $\mu^+$, $\pi^-$, $K^+$ and $K^-$ and can be split up as $A_D =  A(K^+K^-) + A(\mu^+\pi^-)$. Detection asymmetries arise from the fact that the detector is not perfectly symmetric and that it can have different interaction rates for positive and negative particles. They are dependent on the kinematics of the particles and are partially cancelled by periodically reversing the dipole spectrometer magnet polarity. By restricting to the $\phi \rightarrow K^+ K^-$ resonance, the kaon pair is kinematically symmetric, and the measurement is insensitive to any momentum dependent kaon detection asymmetry. The largest systematic uncertainties of the measurement hence come from the tracking and identification asymmetries of the muons and pions which also partially cancel out due to their opposite charge, see table \ref{table:aslssyst}. The remaining tracking asymmetry is determined by using $D^{*+} \rightarrow \Dnul (\rightarrow K^- \pi^+ \pi^+ \pi^-) \pi^-$. Figure \ref{fig:aslsresult} shows a mass fit to the $K^+K^-\pi^+$ invariant mass, where the $D_s^+$ peak is at the right side of the plot; at the left is the $D^+$ peak. The final result from the $\asls = (-0.06 \pm 0.50_\text{stat} \pm 0.36_\text{syst})\%$.

The \asld analysis from \lhcb \cite{asldlhcb} is based on the 2011 and 2012 ($\sqrt{s} = 8$ TeV, $\mathcal{L} = 2 \invfb$) datasets. It looks at the decay of $\Bz \rightarrow D^- \mu \nu_{\mu}$ and  $\Bz \rightarrow D^{*-} \mu \nu_{\mu}$, where the $D^-$ decays to $K^+ \pi^- \pi^-$ and the $D^{*-}$ to $\Dbarnul (\rightarrow K^+ \pi^-) \pi^-$. The measurement uses a time-dependent fit to disentangle $A_P$ from \asld. The fit for the charge asymmetry as a function of decay time for the $D^- \mu^+$ sample is shown in figure \ref{fig:asldresult}. Also for the \asld measurement, the dominating systematic uncertainty comes from detection asymmetries of the final state particles, which is determined using similar methods as for the \asls measurement. The measured value of \asld from \lhcb is $\asld = (-0.02 \pm 0.19_\text{stat} \pm 0.30_\text{syst})\%$ and is, just as the value for \asls, the most precise measurement to date. An overview of all current measurements of \asl including the latest \lhcb results is shown in figure \ref{fig:overview}.

\begin{table}
\begin{center}
\begin{tabular}{l r}
Source of Uncertainty 				&	$\sigma(A_{\mathrm{meas}}) \, [\%]$ \\
\hline
$A(\mu^+ \pi^-)$					& 	$ 0.31 $ \\
$A(K^+K^-)$						& 	$ <0.02 $ \\
Fitting							& 	$ 0.15 $ \\
Background						& 	$ 0.10 $ \\
\hline
Total								&	$ 0.36 $ 
\end{tabular}
\end{center}
\caption{Systematic uncertainties of 2011 \asls analysis.}
\label{table:aslssyst}
\end{table}

\section{Improvements for the 2012 \asls analysis}
\begin{figure}[ht]
\centering
\begin{minipage}[b]{0.47\textwidth}
\includegraphics[width=1.04\textwidth]{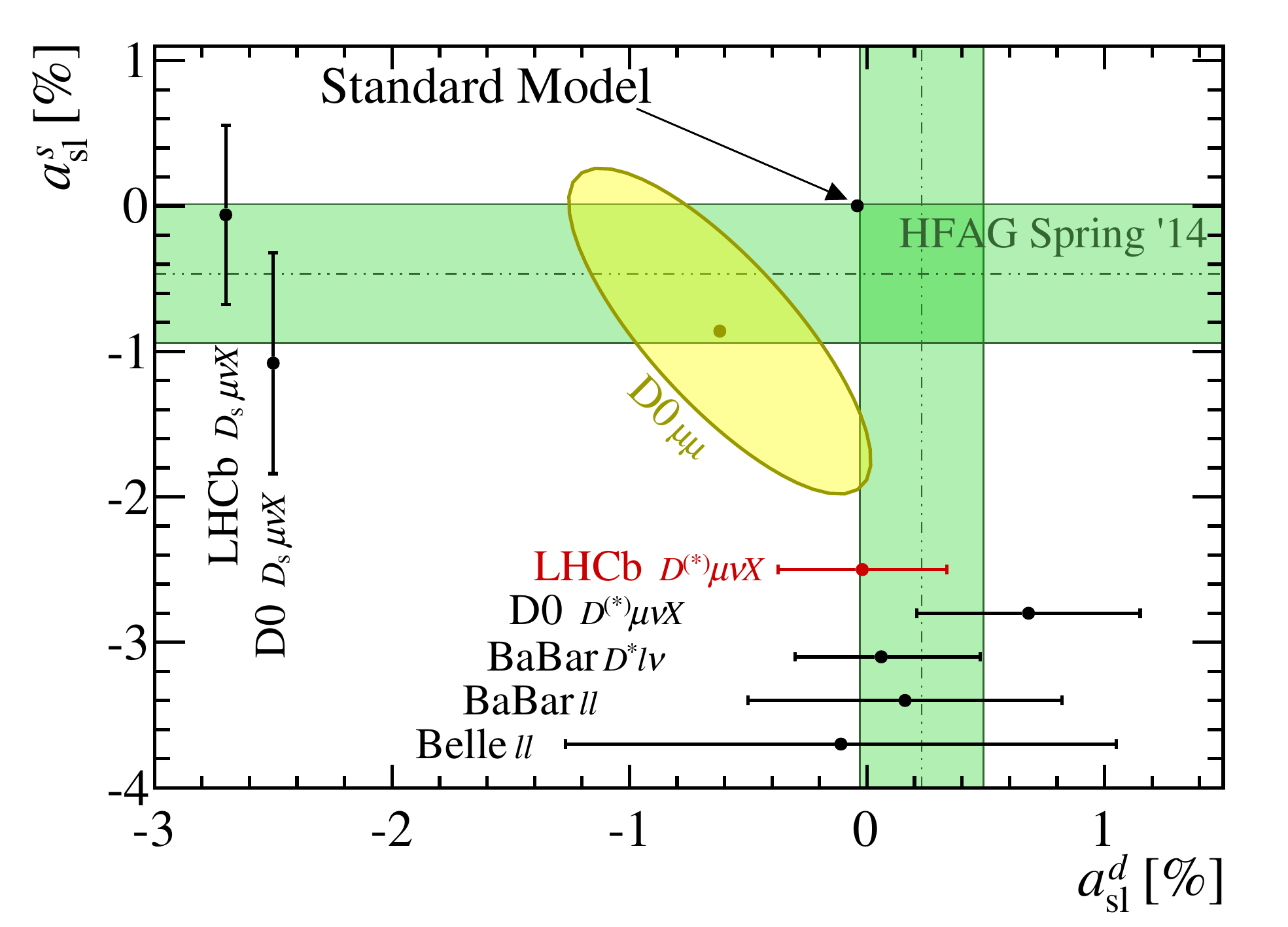}
\caption{Overview of the measurement results of \asls versus \asld. The green bands are the average values of the pure \asls and \asld measurement, excluding the \dzero dimuon measurement, which is shown as the yellow ellipse.}
\label{fig:overview}
\end{minipage}
\quad
\begin{minipage}[b]{0.49\textwidth}
\includegraphics[width=0.85\textwidth]{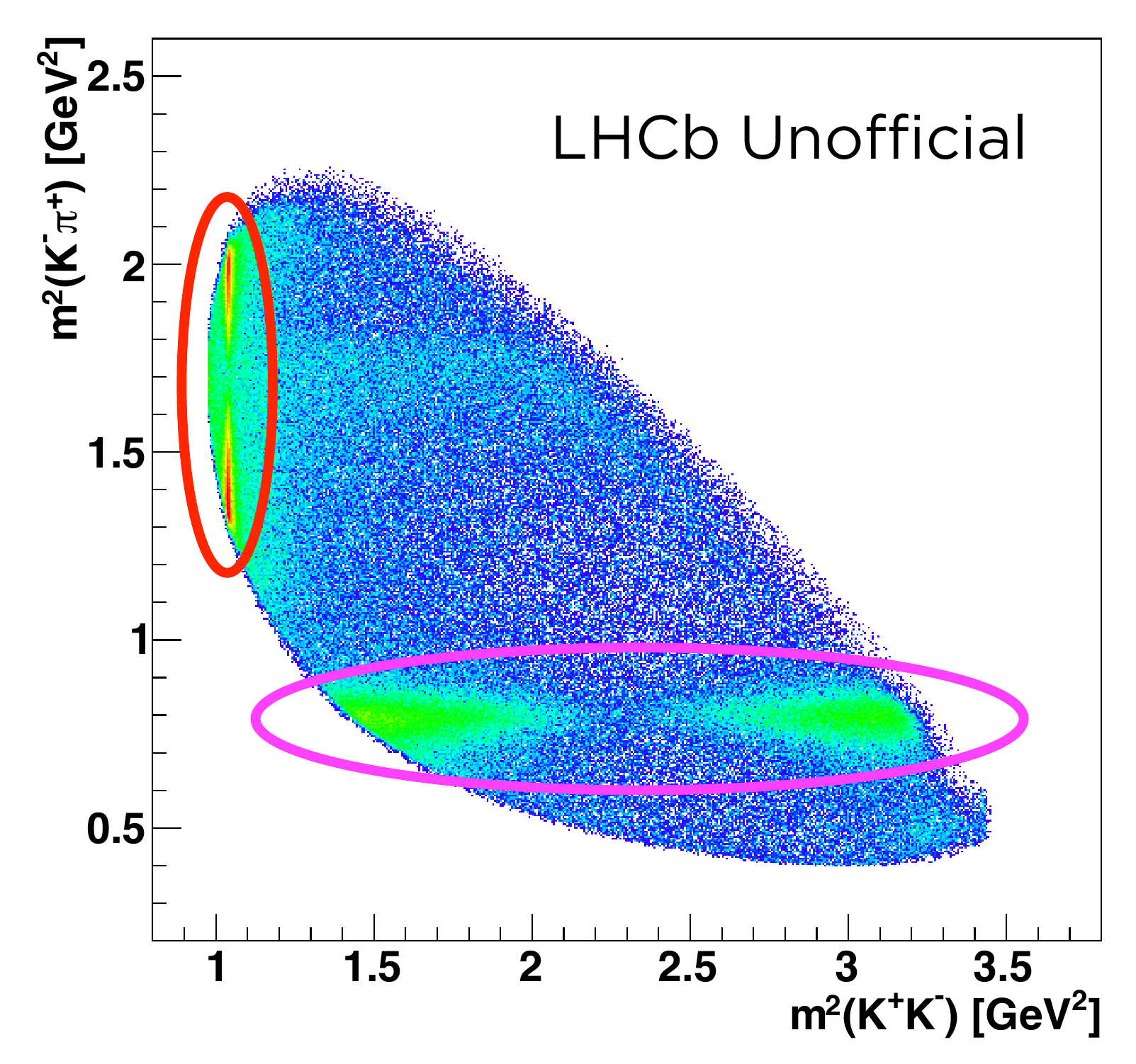}
\caption{The Dalitz plane for $D_s^-  \rightarrow K^+ K^- \pi^-$, where $D_s^-$ comes from $\Bs \rightarrow D_s^-  \mu^+ \nu_{\mu}$. Indicated in red is $D_s^-  \rightarrow \phi (\rightarrow K^+ K^-) \pi^-$, in magenta  $D_s^-  \rightarrow K^{*0} (\rightarrow K^+ \pi^-) K^-$. The remaining phase space will also be analysed}
\label{fig:dalitz}
\end{minipage}
\end{figure}

\noindent For the upcoming analysis on the \asls measurement for \lhcb, the improvements can be divided into two categories: reducing the statistical and systematic uncertainties. The integrated luminosity will be increased by a factor three, reducing the statistical uncertainty. The increase of integrated luminosity also improves the knowledge of the detection asymmetry, which is limited by the size of the control samples. For the 2011 analysis, only the decays of $D_s^-  \rightarrow \phi (\rightarrow K^+ K^-) \pi^-$ were considered. For the future analysis, also $D_s^-  \rightarrow K^{*0} (\rightarrow K^+ \pi^-) K^-$ will be included, as well as the remaining phase space as is indicated in figure \ref{fig:dalitz}.

Different studies are ongoing to improve the tracking and detection asymmetries using new methods. One of them uses prompt charm decays to determine both the muon and pion detection asymmetries. The decays that are used are $D^0 \rightarrow K^- \mu^+ \nu_{\mu}$ and $D^0 \rightarrow K^- \pi^+$. Both are flavour-tagged by requiring them to come from $D^{*+} \rightarrow D^0 \pi^+_s$. The measured asymmetries of these decays can be written as a sum of various production and detection asymmetries, as is done in equation \ref{eq:asymmetries}. Both decays introduce production asymmetries for the $D^{*-}$ and $\pi_s^-$, which cancel out it between the two decay modes. This is also the case for the introduced detection asymmetry of the $K^+$. 
\begin{align}
A_{\text{meas}}(K^+ \mu^-) &= A_D (K^+) + A_D (\mu^-) + A_P(D^{*-}) + A_D (\pi_s^-) \nonumber \\ 
A_{\text{meas}}(K^+ \pi^-) &= A_D (K^+) + A_D (\pi^-) + A_P(D^{*-}) + A_D (\pi_s^-)  \nonumber \\
A_{\text{meas}}(K^+ \pi^-) - A_{\text{meas}}(K^+ \mu^-) &= A_D (\pi^-) +  A_D (\mu^+)
 \label{eq:asymmetries} 
\end{align}

Since the detection and production asymmetries are dependent on the kinematics of the measured particle, asymmetries only cancel when their kinematic distributions are equalised. This has to be done not only for the $D^{*-}$ and $K^+$ particles between the two control modes, but also for the $\mu^-$ and $\pi^+$ particles between the control modes and signal mode. 
\newpage
\noindent In order to equalise the kinematics of all particles, three reweighting steps are defined:
\begin{enumerate}
\item Weight $D^{*+} \rightarrow D^0 (\rightarrow K^- \bm{\pi^{+}}) \pi_s^+$ to $\bar{B}_s^0 \rightarrow D_s^+ (\rightarrow K^+ K^- \bm{\pi^{+}}) \mu^- \bar{\nu}_{\mu}$
\item Weight $\bm{D^{*+}} \rightarrow D^0 (\rightarrow \bm{K^{-}} \mu^+ \nu_{\mu}) \pi_s^+$ to $\bm{D^{*+}} \rightarrow D^0 (\rightarrow \bm{K^{-}} \pi^+) \pi_s^+$
\item Weight $\bar{B}_s^0 \rightarrow D_s^+ (\rightarrow K^+ K^- \pi^+) \bm{\mu^{-}} \bar{\nu}_{\mu}$ to $D^{*-} \rightarrow D^0 (\rightarrow K^+ \bm{\mu^{-}} \bar{\nu}_{\mu}) \pi_s^-$.
\end{enumerate}
After this, mass fits are produced to extract the number of signal events for each of the samples. The reweighting step introduces a change in the effective number of events, $N_{\text{eff}}$, which is defined as $N_{\text{eff}} = \left( \Sigma_{i=1}^N w_i \right)^2  / \Sigma_{i=1}^N w_i^2 $. In this equation, $N$ is the number of candidates, and $w_i$ is the weight of each candidate $i$. The effective number of signal events is calculated and the final asymmetry that is determined includes not only the tracking asymmetries of the pion and muon, but also the muon identification asymmetry and the interaction asymmetry of the pion with the detector: $A_{\mu \pi} = A_{\mu} ^{\text{ID}} + A_{\mu}^{\text{track}} + A_{\pi}^{\text{track}} +  A_{\pi}^{\text{interaction}}$.

\section{Conclusion and Outlook}
\noindent The semileptonic asymmetry \asl provides a method to measure \CP violation in mixing. The \dzero experiment has found values for this asymmetry which are inconsistent with the Standard Model. The \lhcb experiment measured a value for $\asls = (-0.06 \pm 0.50_\text{stat} \pm 0.36_\text{syst})\%$, using $\mathcal{L} = 1.0 \invfb$ and a value for $\asld = (-0.02 \pm 0.19_\text{stat} \pm 0.30_\text{syst})\%$ using $\mathcal{L} = 3 \invfb$. These values do not exclude either the Standard Model or the \dzero values. Using their full 3 \invfb dataset and new methods for determining charge asymmetries, \lhcb will measure a more precise value for \asls. In 2015, the LHC will start running again and with the increased energies, the dataset will be expanded faster, such that by the end of Run-II, a precision better than $10^{-3}$ will be reached. This will allow making a firmer statement on the \dzero measurement as well as probing new physics contributions from beyond the SM.

\bibliographystyle{is-unsrt}
\bibliography{ref}

\end{document}